\documentclass{PoS}

\title{Producing baryons from neutralinos in small H2 clumps over cosmological ages}

\ShortTitle{H2 clumps and WIMPs}

\author{\speaker{Edmond Giraud}\\
        Laboratoire Univers et Particules,
              UMR5299 CNRS-In2p3/Montpellier University, F-34095 Montpellier\\
        E-mail: \email{edmond.giraud@univ-montp2.fr}}

\abstract{Extreme scattering events in quasars suggest  the existence of dark $\rm H_2 ~$ clumps of mass $\rm \sim 10^{-3} ~M_\odot$ and size $\rm \sim 10 ~AU$.  Such  clumps are extremely dense compared to WIMPs clumps of the same mass obtained by N-body simulations.  A WIMP clump gravitationally attracted by a central $\rm H_2 ~$ clump would experience a first infall during which its density increases by $\rm 10^6$ in $\rm \sim 1~ Myr$. In this poster I begin to explore the phenomenology of mixed clumps made with $\rm H_2 ~$ and WIMPs. Molecular clouds built with clumps are efficient machines to transform smooth distributions of WIMPs into concentrated networks. If WIMPs are neutralinos gravitationally attracted in clumps of such molecular clouds, they may either enrich  the baryon sector over cosmological ages, or remain mixed with cold $\rm H_2 ~$ clouds until the clumps evaporate either by collision or by stellar UV heating. A phenomenological model based on an hypothetic dark baryonic component (DBC) that was invoked in the past to explain one of the main drawbacks of CDM profiles, their overly dense cores, is briefly revisited in this context. The DBC is replaced by a mix of $\rm H_2 ~$ and WIMPs, with a small fraction of HI produced by internal $\rm H_2 ~$ collisions,  in slightly dispersive clumpy clouds  that may migrate from the halo towards  inner parts of a galaxy and disk.}

\FullConference{XII International Symposium on Nuclei in the Cosmos,\\
		August 5-12, 2012\\
		Cairns, Australia}
 
\begin{document}

\section{Introduction}
\vspace{-0.2cm}
Most cosmological models studied today are variants of
 a scenario in which perturbations start in a medium of massive weakly interactive particles (WIMPs).
 The present-day universe contains a fraction of non-baryonic dark matter (DM), with a relic density of 
 the order $\rm \Omega_{DM} h_0^2 \approx 0.22$.  If this 
 WIMP dark matter is made of supersymmetric neutralinos, their annihilation mainly produces $\rm \gamma$, protons, antiprotons, electrons 
 and positrons. The annihilation signal from WIMPs in dense substructures of the Milky Way is expected to be 
 stronger than in smooth regions. This question was studied for DM clumps derived from numerical
 simulations, and lead to the conclusion of a large cosmic variance in annihilation products [1] [2].  In this poster I begin to consider the issue of neutralinos gravitationally attracted into clumps of molecular clouds, possibly into dark molecular clouds.  
 A phenomenological model based on an hypothetic dark baryonic component (DBC) that was invoked in the past [28], to explain one of the main drawbacks of CDM profiles, their overly dense cores [3], is briefly revisited in this context. The DBC is replaced by a mix of $\rm H_2 ~$  and WIMPs, with a small fraction of HI produced by internal $\rm H_2 $ collisions, in slightly dispersive clumpy clouds  that may migrate from the halo towards the inner parts of the galaxy and the disk.
\vspace{-0.1cm}
\section{$\bf H_2 ~$ molecular clumps and clouds}
\vspace{-0.2cm}
The estimated value for the stars and gas fraction in disks, $\Omega_{\rm disks} \approx 0.004 \pm 0.002$ [4], compared with the primordial nucleosynthesis requirement, gives room for a dark baryonic component. The existence of small dark clumps of gas in the Galaxy was invoked to
explain Extreme scattering events of quasars [5] [6].
The duration of such events suggests sizes and masses of individual cloudlets of the order ~10 AU and
~$\rm 10^{-3}~M_\odot$. These values of mass and size are also allowed by the searches for lensing toward the LMC [7]. 
Primordial gas consists of atomic hydrogen HI, but in high density cloudlets three-body reactions convert HI to molecular $\rm H_2 $.
Cloudlets can survive in a molecular cloud if the number of collisions is sufficiently small which, according to [8], requires a fractal dimension $1 \leq D \leq 2$. If $D \geq 2.33$, internal and external collisions are too numerous for maintaining a fragmented cloud.

The fractal structure of molecular gas clouds, 
their formation, thermodynamics and stability were intensively studied [9] [10] (most references in [11]). A large fraction of molecular gas in the Galaxy is under the form of  $\rm H_2 $, and cold $\rm H_2 $ may be a significant component of dark matter [12].

\rm
$\rm H_2 $ clumps are extremely dense compared with WIMP clumps of the same mass, as predicted by N-body simulations [13] [14]. More precisely the simulations suggest that a WIMP clump of ~$\rm 10^{-3}~M_\odot$  has a typical size of $\rm \sim 1000~ AU$. Therefore a low density WIMP clump that would be at rest with an $\rm H_2 $ clump would experience a local gravitational infall. In a halo, a smooth distribution of WIMPs, approximately at rest with a network of $\rm H_2 $ clumps, forming a fractal $\rm H_2 $ cloud, will experience as many  local 
infalls as there are $\rm H_2 $ clumps. When WIMPs enter an $\rm H_2 ~$ clump, some of their gravitational energy and momentum are transferred to $\rm H_2 ~$ particles. Some of the low speed WIMPs form a mini-halo. The energy transferred to $\rm H_2 ~$ particles increases their velocity dispersion and therefore the number of $\rm H_2 $ collisions. Dissipative processes start to be at work in the  clump of $\rm H_2 ~$ enriched with WIMPs, including the transform $\rm H_2 ~$ --> HI.
\vspace{-0.2cm}
\section{Baryon factories? Icebergs? }
\vspace{-0.2cm}
When the hydrogen clumps first form, the gas is almost entirely in atomic form. On a time scale of $\rm \sim 10^5~yr$ however, most of the hydrogen is converted to $\rm H_2 $ [8] [10]. Let consider the simple case of a $\rm H_2 $ clump at rest in a WIMP clump of the same mass $\rm 10^{-3}~M_\odot$  The WIMP clump alone would have a radius of 1 000 AU in the absence of $\rm H_2 $. Under the attraction of the central $\rm H_2 $ clump, it will experience a first infall in about 1 Myr during which its density will increase by $10^6$. If WIMPs are typical low mass neutralinos of the order of $\rm m_\chi = 10~ GeV$, number density $\rm n_\chi$, and cross-section $\rm \sigma_{V=0} = 10^{-26} ~cm^3~ s^{-1}$ for annihilation [15], then the WIMPs concentrated into a clump enter a regime of significant annihilation rate, $\rm (1/2 n_\chi )^2 \sigma _{V=0} / t$, leading to a mass transform of $\rm 1.4~ 10^{-7} ~M_\odot ~ Myr^{-1}$. The cumulated mass loss of those neutralinos is 50\% in $\rm \sim 6~ Gyr$. This case may be called a "baryon factory". 

If the neutralino mass is  $\rm m_\chi \sim 100~ GeV~ (see~ Note^1)$, with a cross section $\rm  \sigma_{V=0} = 10^{-27} ~cm^3~ s^{-1}$, the annihilation rate is divided by 1000, and the $\rm H_2 $ clump may build up a WIMP halo over cosmological ages, which might be an "iceberg" in the case of sufficiently high background density. The scale of clumps, is obtained by setting the
free-fall time equals to the Kelvin-Helmholtz time and assuming virialization. The main relation between parameters is that warmer clumps are smaller and more massive, and in that case the number of internal collisions increases. The WIMP mass may smoothly modify the clump equilibrium up to a limit where too large masses become Jeans unstable, fragment and cool.
For a WIMP density similar to that of the solar neighborhood ($\rm 0.3 ~GeV~cm^{-3}$), the amount of neutralinos attracted by a clump in 15-20 Myr is of the order of $\rm 10^{-7} ~M_\odot$ only, namely a small fraction of the clump mass, and a low final density. Higher concentrations require significantly longer infall time.

A molecular cloud in equilibrium with scale 30 pc would have a typical mass $\rm \sim 5 \times 10^5 ~M_\odot$ [16], a density $\rm 18.5~M_\odot ~pc^{-3}$, and a linear separation between clump components $\rm \sim 7 \times 10^{3} ~AU$. If such a cloud  were embedded in a diffuse background of WIMPs, any of the zero velocity WIMP particles would reach an $\rm H_2 $ clump in less than
16 Myr. Therefore the smooth WIMP distribution would quickly become a clumpy network. 

Note$^1$: \sl A weak signal of a broad $\gamma$ line at $\sl E_\gamma \approx 130~ GeV$ was found \rm [17] [31] \sl in Fermi-LAT data \rm [18] \sl near the Galactic Center. It may be interpreted either as a single line or a pair compatible with a $ 127.3 \pm  2.7 ~ GeV$ WIMP annihilating through $\gamma \gamma$ and $\gamma Z$  \rm [19]. \it According to \rm [20] \it an off-centered source at $> 100~GeV$ is associated with the Sgr C molecular cloud.
\vspace{-0.3cm}
\section{ \sl Pamela \bf excess of positrons}
\vspace{-0.2cm}
\rm A rising positron fraction in cosmic rays was discoverred [21] in \it Pamela \rm satellite [22] data.
It was quickly understood that explaining such an effect by sub-products of neutralino annihilations would require a significant boost factor. Numerous and dense halo substructures were invoked to produce such an enhancement. Large dark molecular clouds at rest in a WIMP medium would indeed provide local boost factors. Nevertheless, even  if that question  were resolved, the positron excess would still  be hardly  explained by neutralino annihilations alone, because too many anti-protons would be generated at the same time (references in [23]). Alternately a population of nearby astrophysical sources  (pulsars) would qualitatively explain both the positron and anti-protons spectra  [24], with large uncertainties. So there is still the possibility of a mixed explanation for the \it Pamela \rm positron excess, but  with a large variance and poor constraints.
\vspace{-0.2cm}
\section{Dwarf spiral galaxy profiles}
\vspace{-0.1cm}
That a significant fraction of dark matter may be baryonic (BDM) is suggested by the proportionality between the radial distributions of dark matter and H I gas in a number of faint objects, sometimes all the way through a galaxy [25], and by correlated local density variations between gas and dark matter [26].  Systematic studies suggested that dark matter profiles may be decomposed into two components: one with a distribution similar to that of the H I disk at $\rm r \leq 1.5~r_{\rm opt}$  , and a halo component with a distribution coupled by a tight relation with the disk baryonic matter.
Modeling suggested a systematic increase of the BDM fraction with decreasing luminosity [27], and a larger radial scale length for the BDM than for the HI gas [28].

Let suppose that a fraction of DM is in a form of cold fractal clouds, made with primordial $\rm H_2 $ and gravitationally attracted WIMPs. Far from sources of heating, the gas is thermalized at 3 K, but there are regimes where the cloudlets collide. HI is produced when cloudlets annihilate through collisions. In a collision process the leading parameter is the square of the number density. Then the neutral hydrogen collapses towards the galaxy disk and inner regions, taking some of the bounded WIMPs, and some of the $\rm H_2 $ that was not converted to HI. \it The HI - $ H_2 $  mix is weakly dissipative and should form a flattened BDM halo. \rm Systematic relations between the HI disk, the BDM, and the dark halo, in gas rich spirals of intermediate size, were interpreted as being due to 1) the collisional nature of the $\rm H_2 $ clump annihilations in the outer regions, 2) the dissipative collapse of the diffuse neutral hydrogen towards the disk and intermediate regions from the fractal cold gas in the outer regions, and 3) stellar UV heating in the inner regions (Figure 1). 
\vspace{-4.3cm}
\begin{figure}[htbp]
\begin{center}
\includegraphics[width=10cm]{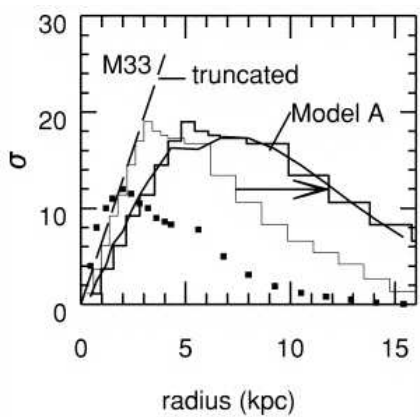}
\vspace{-4.8cm}
\caption{Surface density distributions in M33 from  [29] showing the HI gas (squares), the BDM profile (thick line) and how the
BDM profile can be deduced from the present-day HI distribution. The arrow illustrates the inverse motion of the dissipative gas with respect to the BDM. The dashed line is the inner cut off due to UV heating. If the BDM component is made of H2 clumps it may contain a  WIMP fraction.}
\label{default}
\end{center}
\end{figure}

In dwarf spiral galaxies, the BDM fraction is very large compared with the detected disk component, and the density of DM is low compared with that in bright galaxy halos. All processes involving density are slowed down relative to bright spiral galaxies, but low surface brightness dwarf spirals are particularly interesting for studying links between WIMPs, BDM and HI distributions and processes. The BDM is mostly located at a radius where it makes significant changes to the value of the concentration parameter $c_M$. Its density distribution usually shows a hole in the central region which can be easily understood: if this component is made of molecular hydrogen in a frozen form, heating by the stellar UV flux annihilates the smallest cloudlets, building a cavity in the inner region from which H I evaporates in a diffuse form. In this process the leading parameter is the number density. Those WIMPs that were once bound into dark molecular clouds, may then show density variations following roughly the distribution of evaporated hydrogen. A concordance model for dwarf spiral galaxies was achieved with roughly half the mass in baryons and a mass concentration  $c_M \approx 6.2$ [30], a large fraction which may now be corrected for including both bound and released WIMPs (work in preparation).
\vspace{-0.1cm}
\section{A comment in conclusion}
\vspace{-0.2cm}
I conclude with a remark on the distribution of the double $\rm \gamma$ - ray lines from unassociated Fermi-LAT sources [32], interpreted as being due to neutralino annihilations in sub-halos. These sources were all found within $30^o$ of the galactic plane, while WIMP sub-halos are expected to populate randomly the Galactic halo. If the detections are confirmed, and really due to WIMP annihilations, this may mean that 1) the transport mechanism of WIMPs by $\rm H_2 $ clouds with dissipative HI, and 2) a boost factor by structural overdensities in $\rm H_2 $ cloud  networks, may be meaningful.

\medskip

\bf Acknowledgments: \rm I would like to thank the organizers for the warm and friendly atmosphere of this inspiring conference.

\end{document}